\documentclass [12pt,preprint]{aastex}
\usepackage{epsfig}
\usepackage{natbib}

\begin{document}
\voffset-0.5cm
\newcommand{\gsim}{\hbox{\rlap{$^>$}$_\sim$}}
\newcommand{\lsim}{\hbox{\rlap{$^<$}$_\sim$}}

\title{Are Superluminous Supernovae  Powered By Collision Or\\ 
             By Millisecond Magnetars?}

\author{Shlomo Dado\altaffilmark{1} and Arnon Dar\altaffilmark{1}}

\altaffiltext{1}{Physics Department, Technion, Haifa 32000, Israel}

\begin{abstract} 

Using our previously derived simple analytic expression for the bolometric 
light curves of supernovae, we demonstrate that the collision of the fast 
debris of ordinary supernova explosions with relatively slow-moving shells 
from pre-supernova eruptions can produce the observed bolometric light 
curves of superluminous supernovae (SLSNe) of all types. These include 
both, those which can be explained as powered by spinning-down millisecond 
magnetars and those which cannot. That and the observed close similarity 
between the bolometric light-curves of SLSNe and ordinary interacting SNe 
suggest that SLSNe are powered mainly by collisions with relatively slow 
moving circumstellar shells from pre-supernova eruptions rather than by 
the spin-down of millisecond magnetars born in core collapse supernova 
explosions.

\end{abstract}

\keywords{supernovae: general}

\maketitle

\section{Introduction}

For a long time, it has been widely believed that the observed luminosity 
of all known types of supernova (SN) explosions is powered by one or more of 
the following sources: radioactive decay of freshly-synthesized elements, 
typically $^{56}$Ni (Colgate and McKee~1969;  Colgate et al.~1980), heat 
deposited in the envelope of a supergiant star by the explosion shock 
(Grassberg et al.~1971), and interaction between the SN debris and the 
circumstellar wind environment (Chevalier 1982). Recently, however, a new 
type of SNe, superluminous SNe (SLSNe) whose peak luminosity exceeds 
$10^{44}$ ergs per second, much brighter than that of the brightest normal 
thermonuclear supernovae (type Ia) and core collapse supernova (types Ib/c 
and II) was discovered by modern supernova surveys. The first discovered 
SLSNe was SN2006gy (Quimby 2006; Quimby et al.~2007) in the Texas 
Supernova Search (Quimby et al.~2005) and followed  by 
Smith et al.~(2007,2008) and by Ofek et al.~(2007). A few more SLSNe were 
discovered in the Texas Supernova Search and many more in deeper and wider 
surveys with the Palomar Transient Factory (PTF, Rau et al. 2009), the 
Panoramic Survey Telescope \& Rapid Response System (Pan-STARRS, Kaiser et 
al. 2010), the Catalina Real-time Transient Survey (CRTS, Drake et al. 
2009), and the La Silla QUEST survey (Hadjiyska et al. 2012). In analogy 
to the spectroscopic classification of ordinary SNe, Gal-Yam (2012) has 
recently classified SLSNe into two distinct groups, Hydrogen-rich events 
(SLSN Type II) and events lacking spectroscopic signatures of hydrogen 
(SLSNe Type I), which was further divided into a minority of events 
whose luminosity appears to be dominated by radioactivity (SLSN-R) and a 
majority that require some other source of luminosity (SLSN-I).

Following the discovery of SN2006gy, Woosley, Blinnikov and Heger (2007) 
suggested that it is a pair-instability supernova explosion (Rakavy and 
Shaviv~1967; Barkat et al.~1967) of a very massive star whose luminosity 
was powered by radioactive $^{56}$Ni (see also, Kasen et al.~2011) 
synthesized in the SN explosion.  However, the peak luminosity of 
SN2006gy, ${\rm L_p\approx 3\times 10^{44}\, erg\, s^{-1}}$ at peak-time 
${\rm t_p\approx 70\,d}$ (Smith et al.~2007; Ofek et al.~2007), would have 
required the synthesis of more than ${\rm 40\, M_\odot}$ of $^{56}$Ni! 
(See Eq.~(6)), which is highly implausible.  Consequently, Smith and 
McCray (2007) suggested that the light-curve of SN200gy was produced by an 
SN blast wave that breaks free of an opaque circumstellar shell into a 
surrounding wind.

An alternative power-source of the luminosity of SLSNe that has been 
suggested more recently is the spin-down of a millisecond magnetar born in 
SN explosions of massive stars (Kasen \& Bildsten~2010; 
Woosley~2010).\footnote{Ordinary magnetars are slowly rotating 
neutron stars with observed periods P$\sim 2-12$ s and ${\rm\dot{P}\sim 
10^{-13}-10^{-10}}$, whose internal magnetic fields are believed to be 
${\rm B\sim 10^{14}-10^{15}}$ Gauss in excess of the quantum critical 
value ${\rm B_{QED}=4.4\times 10^{13}}$ Gauss. Their huge magnetic field 
energy was inferred from the assumption that their observed spin-down rate 
is due to magnetic dipole radiation.}. Magnetars were suggested before to 
power 
soft gamma repeaters (Katz~1982; Thompson and Duncan~1995; Kouveliotou et 
al. 1998), anomalous X-ray pulsars (Thompson and Duncan~1996) and gamma 
ray bursts (Usov 1992; Zhang and Meszaros~2001). However, despite their 
popularity, there are unresolved problems with magnetar models (e.g., Katz 
2013), which were realized long ago and led to alternative 
conventional explanations for the rapid spin down of slowly rotating 
neutron stars,  their energy source, and the observed properties of 
soft gamma-ray repeaters (SGRs), anomalous X-ray pulsars (AXPs),
and gamma 
ray bursts (GRBs). E.g., rapid braking of neutron stars can be due to 
emission of relativistic particles/wind/jets along their open magnetic 
field lines (e.g., Marsden et al.~1999;  Dar and De R\'ujula 2000a) rather 
than by magnetic dipole radiation. Their bursts and steady state emission 
can be powered by a sudden or gradual local or global phase 
transition/contraction to a more condensed state (Dar and De R\'ujula 
2000a). Moreover, recent measurements of the period derivatives of SGRs 
0418+5729  and 1822-1606 (Rea et al.~2010,2012, respectively), and 3XMM 
J185246.6+003317 (Rea et al.~2013), imply that their dipole magnetic 
fields are well in the range of ordinary radio pulsars and challenge the 
magnetar model of SGRs and AXPs. In fact, magnetars are needed to 
explain the observed properties neither of SGRs and AXPs nor of GRBs, and 
may just be a fiction.

Long duration GRBs and their afterglow are very well explained by the 
cannonball (CB) model of GRBs (e.g., Dar \& R\'ujula 2004; Dado et 
al.~2009 and references therein). In the CB model, long duration GRBs are 
produced by inverse Compton scattering of glory\footnote{The glory is a 
light halo around the progenitor star, a Wolf Rayet (WR)  or large blue 
variable (LBV), formed by stellar light scattered from circumstellar 
shells blown from the progenitor star in eruptions prior to its SN 
explosion.} by highly relativistic bipolar jets of plasmoids (cannonballs) 
of ordinary matter, which presumably are ejected in mass accretion 
episodes of fall-back material on the newly formed central object (neutron 
star or black hole) in stripped-envelope supernova explosions. Such a 
circumstellar (cs) surroundings of stripped-envelope core collapse SN 
explosions can explain (Dado and Dar 2012)  the bolometric light-curves of 
both SLSNe-I and SLSNe-II as a plastic collision between the fast debris 
from core collapse SN explosions of types Ic (SNeIc) and IIn (SNeIIn) and 
the slower massive circumstellar shells (Dado and Dar 2012): SLSNe of 
types I and II could simply be ordinary SNeIb/c and SNeIIn, respectively, 
which become superluminous by such collisions. Moreover, Dado and Dar 
(2012,2013) have demonstrated that the bolometric light curves of SNeIa, 
SLSNe-I, SNeIb/c and SLSNe-Ib/c can all be well described by a simple 
analytic expression (a universal master formula) for the light curves of 
supernova explosions powered by an ordinary amount of radioactive 
$^{56}$Ni and dominated by the collision of the SN debris with 
pre-supernova ejecta.

Recently, however, Inserra et al.~(2013) and Nicholl et al.~(2013) have 
demonstrated that using a spinning-down millisecond magnetar as a 
power-source in the master formula derived by Dado and Dar (2013) for the 
bolometric light curves of SN explosions, they could reproduce quite well 
the bolometric light curves of SLSNe-Ic\footnote{Inserra et al.~(2013) 
obtained the master formula that was derived by Dado and Dar~2013, by 
combining Eqs. 31, 32, and 36 of Arnett 1982.}. They concluded that the 
light-curves of all known SLSNe-Ic may be explained by spinning down 
millisecond magnetars. But so far the magnetar model has not been 
demonstrated to be able to explain multi-peak bolometric light-curves or 
fast decline light curves of some SLSNe-Ic as well as SLSNe-II.

In contrast, in this letter we demonstrate that a mass of radioactive 
$^{56}$Ni similar to that synthesized in ordinary SNeIc/b and SNeIIn 
explosions plus collision of the fast SNe debris with 
slowly expanding dense circumstellar 
shells formed by eruptions before the SN explosion can explain well the 
observed bolometric light curves of all types of SLSNe, including those 
with a multipeak structure (two or more peaks) such as SN2006oz (Leloudas 
et al.~2012) and PTF12-dam (Nicholl et al.~2013), or with a sudden fast 
late-time decline such as that of SN2006gy (Smith et al.~2008). Such 
bolometric light-curves that look similar to those of ordinary interacting 
SNeIb/c and SNeIIn, such as SN2005bf (Folatelli et al. 2006), SN2009ip 
(Pastorello et al. 2013: Margutti et al.~2013) and SN2010mc (Ofek et 
al.~2013)  are shown in this letter to be well explained as interacting 
SNe. This suggests that SLSNe types I and II can simply be ordinary 
SNeIb/c and SNeIIn, respectively, which are powered by the decay of an
ordinary amount of $^{56}$Ni and become superluminous mainly by the 
interaction of their fast SN debris with massive circumstellar 
shells\footnote{In many distant SLSNe, where only a modest mass of 
$^{56}$Ni is synthesized in the explosion, the initial SNe may not be 
bright enough to be visible or resolved from the SLSNe light curve.}.
 
\section{The master equation for supernova light curves}
Dado and Dar (2013) have derived a simple master formula 
for the bolometric luminosity ${\rm L_b(t)}$ of SN explosions
approximated by a fireball with homologous expansion and 
photon escape by random walk to its surface,
\begin{equation}
{\rm L_b(t)={e^{-t^2/2\, t_r^2}\over t_r^2}
\int_0^t t\,e^{t^2/2\, t_r^2} \dot{E}\, dt.}
\label{Lbol}
\end{equation}
where t is the time after shock break out, ${\rm \dot{E}(t)}$  is 
the energy deposition rate in the SN fireball,
and the mean escape time of optical photons by diffusion  has the 
approximate time-behaviour,
${\rm t_{dif}=t_r^2/t}$ with a diffusion time scale (the time when 
${\rm t_{dif}=t_r=t}$), 
\begin{equation}
{\rm t_r \approx \left[{3\, M\, f_e\, \sigma_{_T}
      \over 8\,\pi\, c\,V}\right]^{1/2}\,,}
\label{tr}
\end{equation}
where M is the expanding mass, V is its expansion velocity,  
${\rm f_e}$ is the fraction of free (ionized) electrons in M 
and  ${\rm \sigma_{_T}}$  is the Thomson cross section. 

Assuming that the 3s and 3p electrons outside the neon-like closed shells 
core of intermediate mass elements, such as Mg, Si, and S, are ionized, 
one 
obtains ${\rm f_e(Si)\approx 0.285}$ whereas, e.g., for cobalt the 4s and 
3d electrons outside the argon-like closed shell core, are ionized 
yielding ${\rm f_e(Co)=0.333}$. For ${f_e\approx 0.31\pm 0.03}$ we expect 
${\rm t_r\approx (9 \pm 1)\,(M/M_{\odot})^{1/2}} $ day.

\section{The power supply in  interacting SNe}
For the sake of simplicity, let us assume that SLSNe are powered by the 
radioactive decay chain ${\rm^{56}Ni\rightarrow ^{56}Co\rightarrow 
^{56}Fe}$, 
which begins at t=0, and by plastic collision that begins 
later at time ${\rm t_c > 0}$ between the expanding fireball and a 
circumstellar 
mass, which was blown off from the progenitor star sometime before the 
explosion and expands with velocity ${\rm V_{cs}\ll V}$. Hence 
${\rm \dot{E}=\dot{E}_{r} +\dot{E}_c}$ and ${\rm L_b(t)=L_r(t)+L_c(t)}$,
where the subscripts r and c denote 
radioactivity and collision, respectively.

\noindent 
{\bf Radioactive Power:}
Energy deposition by Gamma-rays and positrons emitted in the decay chain  
${\rm ^{56}Ni\rightarrow ^{56}Co\rightarrow ^{56}Fe}$
can power SN light-curves at a rate 
${\rm \dot{E}_{r}= \dot{E}_\gamma+ \dot{E}_{e^+}}$ where,
\begin{equation}
{\rm \dot{E_{\gamma}}={\rm {M(^{56}Ni)\over M_\odot}\,[A_\gamma(Ni)\, 
7.85 \times
   10^{43}\,e^{-t/8.76\,d} + A_\gamma(Co)\, 1.43\times 10^{43}\,
   [e^{-t/111.27\,d}-e^{-t/8.76\,d}]]}\,\, {\rm erg\, s^{-1}}\,.}
\label{Pgamma}
\end{equation}
${\rm A_\gamma(Ni)}$ and  ${\rm A_\gamma(Co)}$ are the attenuation 
coefficients  of the ${\rm \gamma}$-rays
from the decay of $^{56}$Ni and $^{56}$Co, respectively, in the SN
fireball.  ${\rm A_\gamma
\approx 1-e^{-\tau_\gamma}}$, where ${\rm \tau_\gamma(t)\approx 
R\,\Sigma_i n_i\sigma_i}$ is the fireball opacity  and the summation
extends over all its
particles. For a composition dominated by intermediate mass elements and 
iron group elements,
${\rho_i\,\lambda_i\approx 13\,  g\, cm^{-2}}$ for the  ${\rm \gamma}$ 
rays
with ${\rm <E_\gamma>\approx 0.54}$ MeV from the decay of $^{56}$Ni
and larger by  a factor $\approx 5/3$ for  ${\rm \gamma}$-rays with
${\rm <E_\gamma>\approx 1.32}$
MeV from the decay of $^{56}$Co (see  Hubbell~1982). Hence,
${\rm \tau_\gamma=R\, \rho /\rho_i\lambda_i\approx t_\gamma^2/t^2}$ where
${\rm t_\gamma^2\approx 3\, M/4\, \pi\, V^2\,\rho_i\,\lambda_i} $, and
${\rm t_\gamma(Ni) \approx 35\,(M/M_\odot)}$ d while  
${\rm t_\gamma(Co)\approx 22\, (M/M_\odot)}$ d.

The positrons from the ${\rm \beta^+}$-decay of $^{56}$Co (and the 
${\rm e^{\pm}}$
from the decay of other relatively long lived radioactive isotopes that
were synthesized in the thermonuclear explosion) are presumably trapped
by the  turbulent magnetic field of the SN fireball. 
Presumably, they lose their 
kinetic energy before the
${\rm e^+e^-}$ annihilation into two 0.511 MeV ${\rm \gamma}$-rays.
Their energy deposition rate is given approximately  by  
\begin{equation}
{\rm \dot{E_{e^\pm}}= {\rm {M(^{56}Ni)\over 
M_\odot}\,A_e\,(1+6.22\,A_\gamma)\,
1.43\times 10^{43}\, [e^{-t/111.27\,d}-e^{-t/8.76\,d}]}\, {\rm erg\, 
s^{-1}}\,.}
 \label{betaplus}
\end{equation}
where ${\rm A_{e}=0.034}$ is the ratio between the energy
released  in the radioactive decay of $^{56}$Co as kinetic energy of
$e^+$ and that released directly in $\gamma$-rays.
The kinetic energy deposition dominates the radioactive power supply   
when the fireball becomes highly transparent to
${\rm \gamma}$-rays, i.e., when ${\rm t\gg t_\gamma(Co)}$.

As long as ${\rm \dot{E}}$ changes rather slowly with time
relative to the fast rise with time of the factor 
${\rm t\,e^{t^2/2\, t_r^2}}$, it can be  
factored out of the integration in Eq.~(1) to yield
\begin{equation}
{\rm L_r(t)\approx [1-e^{-t^2/2\, t_r^2}]\, \dot{E}_{r}\,.}
\label{approxLr}
\end{equation} 
Hence, the contribution of radioactivity to bolometric luminosity
initially rises approximately like ${\rm L_r\approx 
(t^2/2\,t_r^2)\,\dot{E}_{r}}$, has a peak value
${\rm L_r(t_p)\approx \dot{E}_r(t_p)}$,
and a late-time asymptotic behaviour ${\rm L_r\approx \dot{E}_{r}}$.

If the bolometric luminosity of an SLSN is  powered by the radioactive 
decay chain ${\rm ^{56}Ni\rightarrow ^{56}Co\rightarrow ^{56}Fe}$, and if
its luminosity peak time ${\rm t_p \gg \tau(Ni)=8.76\,d}$, then
its peak luminosity satisfies (Dado and Dar 2012)
\begin{equation}
{\rm L(t_p)=\dot{E}_r\approx 1.43\times 10^{43}\,[M(^{56}Ni)/M_\odot]\,
A_\gamma(Co)\, e^{-t_p/111.27\,d}\,\,\,erg\,s^{-1}}\,.
\label{Mdot}
\end{equation}
E.g., if the bolometric luminosity of SN2006gy was powered by 
the synthesis of $^{56}Ni$, then its observed peak value 
${\rm L_p(70d)\approx
3\times 10^{44}\, erg\, s^{-1}}$ would have required 
${\rm M(^{56}Ni)\geq 40 M_\odot}$ ! 

\noindent
{\bf Collision Power:} 
For simplicity, consider  a plastic collision between a 
fast expanding SN fireball and a slowly moving 
circumstellar shell (CS) with  a density profile
${\rm \rho(R)=0}$ for ${\rm R<R_c}$,
${\rm \rho(R)=\rho_0\ R_0^2/R^2}$ for ${\rm R_c\leq R\leq R_e}$, 
and ${\rm \rho(R)=0}$ for ${\rm R>R_e}$.
Such a shell could have been formed by mass ejection at a 
constant rate ${\rm \dot{M}}$ during  an eruption time 
${\rm \Delta t}$ sometime before the SN explosion, yielding 
${\rm \rho_0\ R_0^2 =\dot{M}/4\,\pi\, V_{cs}}$. 
Hence,  for ${\rm t_c\leq t\leq t_e}$, 
and  ${\rm V \gg V_{cs}}$, 
the energy  deposition rate in the SN fireball is
\begin{equation}  
{\rm \dot{E}_c(t)= {\dot{M}\, (V-V_{cs})^3 \over 2\,V_{cs}}
\approx  {\dot{M}\, V^3 \over 2\,V_{cs}}}\,.
\label{Pc}
\end{equation}
The matter swept-up by the expanding SN fireball 
decelerates the SN  expansion. For a relatively low ${\rm V_{cs}}$,
momentum conservation yields for ${\rm t_c\leq t \leq t_e}$
\begin{equation}  
{\rm V(t)\approx V_c\,[1+b\,(t-t_c)]^{-1/2}}\,, 
\label{V}
\end{equation}
where ${\rm V_c=V(t_c)}$ and 
${\rm b\approx 8\,\dot{M}\,V_c/3\,M\,V_{cs}}\,.$

During the collision,  $\dot{E_c}$ changes with time rather slowly 
relative to the fast rise of  the exponential factor in the 
integrand on the right hand side of  Eq.~(1), which can be approximated by 
\begin{equation}
{\rm L_c(t)\approx [1-e^{-(t-t_c)^2/2\, t_r^2}]\, \dot{E}_{c}\approx
 [1-e^{-(t-t_c)^2/2\, t_r^2}]\, \dot{M}\,V^3/2\,V_{cs}}\,.
\label{approxLc}
\end{equation}
Hence, the contribution of a collision with a circumstellar shell/wind to 
the bolometric luminosity  rises approximately like 
${\rm L_c\approx (t-t_c)^2/2\,t_r^2)\,\dot{E}_{c}(t)}$, and has a peak 
value 
${\rm L_c(t_{pc})=\dot{E}_c(t_{pc})}$ at the peak time ${\rm t_{pc}}$
of the collision luminosity
and a late-time asymptotic behaviour 
${\rm L_c(t)\approx \dot{E}_c(t)}$ as long as ${\rm t \leq t_e}$.
Beyond ${\rm t_e}$, where ${\rm \dot {E}_c(t)=0}$, Eq.~(1) yields
\begin{equation}
{\rm L_c(t>t_e)=L_c(t_e)\, exp^{-[(t-t_c)^2-(t_c-t_e)^2]/2\,t_r^2}}. 
\label{Lctgte}
\end{equation}
Note that during the collision, ${\rm [t_r]^2}$ as given by Eq.~(2) 
increases with time roughly like ${\rm M\, V_0/ V^2}$. 

The generalization of the above formulae to SN explosion expanding into a 
pre-supernova wind or colliding with a sequence of circumstellar shells is 
straight forward. The bolometric light-curve during a collision with a 
continuous wind is obtained by setting $t_c$ to be the beginning time of 
the explosion. For expansion into a multi-shell environment, each SN-shell 
collision is still described by Eqs.~(8)-(10) with $t_c$ and 
${\rm t_e=t_c+dt_c}$ being, respectively, the beginning and end times of 
the collision of the expanding SN with the specific shell.

The pre-supernova history of the progenitors of SNe-Ib/c SNeIIn, and 
consequently their environments, are poorly known. The assumed constant 
rate of mass ejection during mass ejection  episodes and their 
assumed sharp step-like beginning and decline are surely over 
simplifications, which were introduced 
in order to minimize the number of adjustable parameters in our           
fitted bolometric light curves of SLSNe dominated by collisions. 
If, however, the SN expansion velocity ${\rm V(t)}$ is known 
from the time-dependent profiles of SN absorption and emission lines, then 
Eq.~(9) may be used to extract roughly the pre-supernova history of mass 
ejection.

\section{Colour temperature for collision dominated lightcurve}
During the photospheric phase, when the SN fireball is optically thick, 
its continuum  spectrum is approximately that of a black body,
with an effective  photospheric temperature that satisfies 
the Stefan-Boltzmann law. Consequently,
\begin{equation}
{\rm T\approx \left[{L_b \over 4\,\pi\, V^2\,t^2\,\sigma}\right]^{1/4}\,} 
\label{Tc}
\end{equation}
where ${\rm \sigma=5.67\times 10^{-5}\, erg\, s^{-1}\, cm^{-2}\, K^{-4}}.$
Thus, if radioactivity can be neglected  during  the  collision, then 
${\rm T \propto [V/2\, t_r^2]^{1/4}}$ 
for ${\rm (t-t_{c})^2 \ll t_r^2}$, while ${\rm T\propto V^{1/4}\, 
(t-t_c)^{-1/2}}$ for ${\rm (t-t_c)^2 \gg t_r^2}$.
The two behaviours can be combined roughly into a
simple interpolation  formula,
\begin{equation}
{\rm T\propto  [V/[2\,t_r^2+(t-t_c)^2)]^{1/4}}\, .
\label{TVc}
\end{equation}
This formulae is valid until the SN  enters the nebular phase. 

\section{Comparison with observations} Figure 1 compares the observed 
bolometric light curve of SN2009ip (Margutti et al.~2013) and that 
expected from its description as an "ordinary" interacting SNIIn.  
Similar fits were obtained for the observed bolometric light-curves of the 
SN2005fb (Folatelli et al.~2006) and SN2010mc (Ofek et al.~2013) when they 
were treated as interacting ordinary SNIb/c and SNIIn, respectively. Figs. 
2-4 compare the bolometric light curves of the superluminous supernovae 
2006oz (Leloudas et al.~2012), PTF 12dam (Nicholl et al.~2013) and 
PS1-11ap (McCrum et al.~2013) and their description as interacting SNIb/c 
or SNIIn. Similar fits were obtained for other typical SLSNe, such as 
SN2006gy (Smith et al.~2008)  and SN2010gx (Pastorello et al. 2010). The 
best fit parameters of all these fits are listed in Table 1. As can be 
seen from Figs. 1-4 and Table I, the bolometric light curves of all the 
above representative SNeIc, SNeIIn, SLSNe-Ic and SLSNe-II are described 
well by our master formula for interacting SNe.

\section{Conclusions}
Models where $^{56}$Ni is the power source for the very large luminosities 
of SLSNe may account for the light curve rise-time and peak-value 
reasonably well. However, the peak luminosities of SLSNe with 
a relatively slow-rising 
light-curve require the synthesis of an implausible amount of 
$^{56}$Ni, e.g., ${\rm M(^{56}Ni)\approx 40 M_\odot}$ in the case of 
SN2006gy. Such models also fail to fit well the late-time decline of the 
bolometric luminosities of several SLSNe.

Millisecond magnetars, if born in SNeIc explosions, were shown to be able 
to produce the observed bolometric lightcurves of SLSNe-Ic (e.g., Inserra 
et al.~2013, Nicholl et al.~2013) reasonably well. However, it remains to 
be seen whether magnetar powered SLSNe-Ic, together with a modest amount 
of 
radioactive $^{56}Ni$ synthesized in the explosions, can also explain 
multipeak light-curves and the observed kinematic and spectroscopic 
properties of SLSNe-Ic. Moreover, despite their popularity, there are 
unresolved problems with magnetar models (e.g., Katz 2013), there are 
observations that challenge the magnetar paradigm (e.g., Rea et al. 
2010,~2012,~2013), and there are more conventional 
explanations for almost all the phenomena that were interepreted 
with magnetar models.

In this letter, we have demonstrated that SLSNe may be 
ordinary SNe of types Ib/c and IIn, which interact with the circumburst 
environment created by their progenitor stars prior to their SN explosion. 
This is suggested by their similarity to interacting SNe of types Ic and 
IIn. Our simple master formula (Eq.~(1)), which was derived in 
Dado and Dar~2013 for the bolometric light-curves of SN explosions 
fit well the 
bolometric light curves of both interacting SNe and SLSNe, with very few 
parameters. This success is despite the over-simplified geometry, 
hydrodynamics and radiation transport used in its derivation: The density, 
internal energy and velocity profiles of the SN fireball resulting from a 
collision of the SN debris with a circumstellar shell are surely more 
complicated than those assumed, and so are the conversion of kinetic 
energy into thermal energy via forward and backward shocks and its 
transport to the SN photosphere from where it escapes freely into space. 
Nevertheless, as demonstrated in this paper, simple modeling of SLSNe as 
interacting SNe can reproduce well the observed variety of their 
bolometric light curves; fast and slowly rising, fast and slowly decaying, 
single and multi-peak light curves. Moreover, the master formula (1) 
together with Eqs. (9) and (11) successfully predict the correlated 
behaviours as a function of time of the observed bolometric luminosity, 
the 
colour temperature and the SN expansion velocity inferred from the shapes 
of absorption and emission lines in several SLSNe, and can be used to 
extract the eruption history (${\rm \dot{M}/V_{cs}}$) of the progenitor 
star 
before its SN explosion (Dado and Dar, in preparation). They imply that 
SLSNe are produced by the explosion of very massive stars, akin to eta 
Carinae, that have lost tens of solar masses in relatively short emission 
episodes during a couple of centuries prior to their final SN explosion.

\begin{deluxetable}{lllllllllll}
\tablewidth{0pt}
\tablecaption{Best fit parameters 
of the bolometric light-curves of a representative sample of interacting 
supernovae}
\tablehead{
\colhead{SN}&\colhead{M($^{56}$Ni)}&\colhead{$t_0$[d]}&\colhead{$t_r$[d]}&
\colhead{$t_c$[d]}&\colhead{$dt_c$[d]}&\colhead{$t_{rc}$[d]}&
\colhead{$\dot{E}_c(t_c)$[erg]}&\colhead{b[d$^{-1}$]}&
\colhead{$\chi^2/dof$}}
\startdata
SN2009ip &0.024$M_\odot$&-5.53&23.0&48.1&14.6&6.7& 3.7E42 &.153 &0.98\cr 
         &                 &   & & 83.8 &4.64 & 4.98&+4.5E41  &.22 &  \cr
         &                 &   & & 104.9&12.2 & 5.02&+3.0E40 &.07  &  \cr
SN2006oz & 0.95$M_\odot$&-10.56&8.28&8.66&36.9&23.1&1.23E44&.035&1.08\cr
PTF12dam& 2.45$M_\odot$  &-66.0 & &-64.1 &70.8 & 62.8&1.34E44&.038&0.29\cr
PS1-11ap & 0.072$M_\odot$ &-48.9&21.4&-30.8&51.5& 19.0&2.32E43&.069&0.17\cr
SN2005bf & 0.147$M_\odot$ &-42.4 & 14.7&-18.8&24.9&12.9&2.1E42&.038&1.34\cr
SN2006gy & 0.19$M_\odot$ &2.7& 5.96& 5.26 &82.7&41.5&8.15E43&.013&0.37\cr
         &                 &   & & 214. &     &    &+1.5E42&.0004& \cr    
SN2010gx & 0.012$M_\odot$&-28.1 & 8.27&-22.5&37.5&10.3&1.04E44 &.06&1.12\cr   
\enddata
\end{deluxetable}

\newpage \begin{figure}[] 
\centering 
\vspace{-2cm} 

\epsfig{file=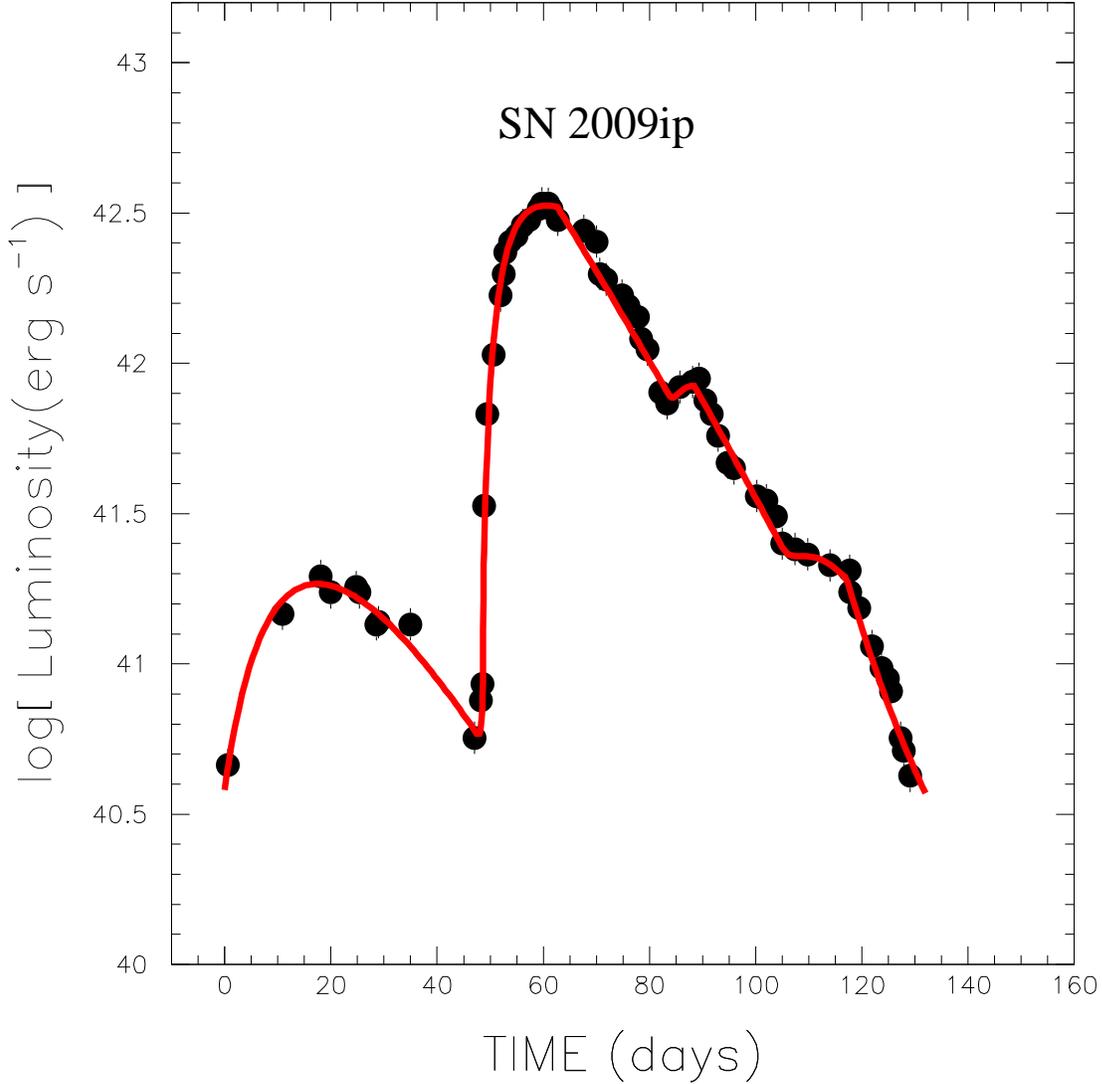,width=16.cm,height=16.cm} 
\caption{Comparison between the rest frame light-curve of the bolometric 
luminosity of supernova 
SN2009ip (Margutti et al.~2013)and a best fit
light-curve of a supernova powered by the decay of 
${\rm 0.025\, M_\odot}$ of 
$^{56}$Ni synthesized in the explosion and subsequent plastic collisions 
of the SN debris with a succession of three  massive shells 
after a wind blown by the  
progenitor star sometimes before its  explosion.}
\label{Fig1} 
\end{figure}

\newpage 
\begin{figure}[]
\centering
\vspace{-2cm}
\epsfig{file=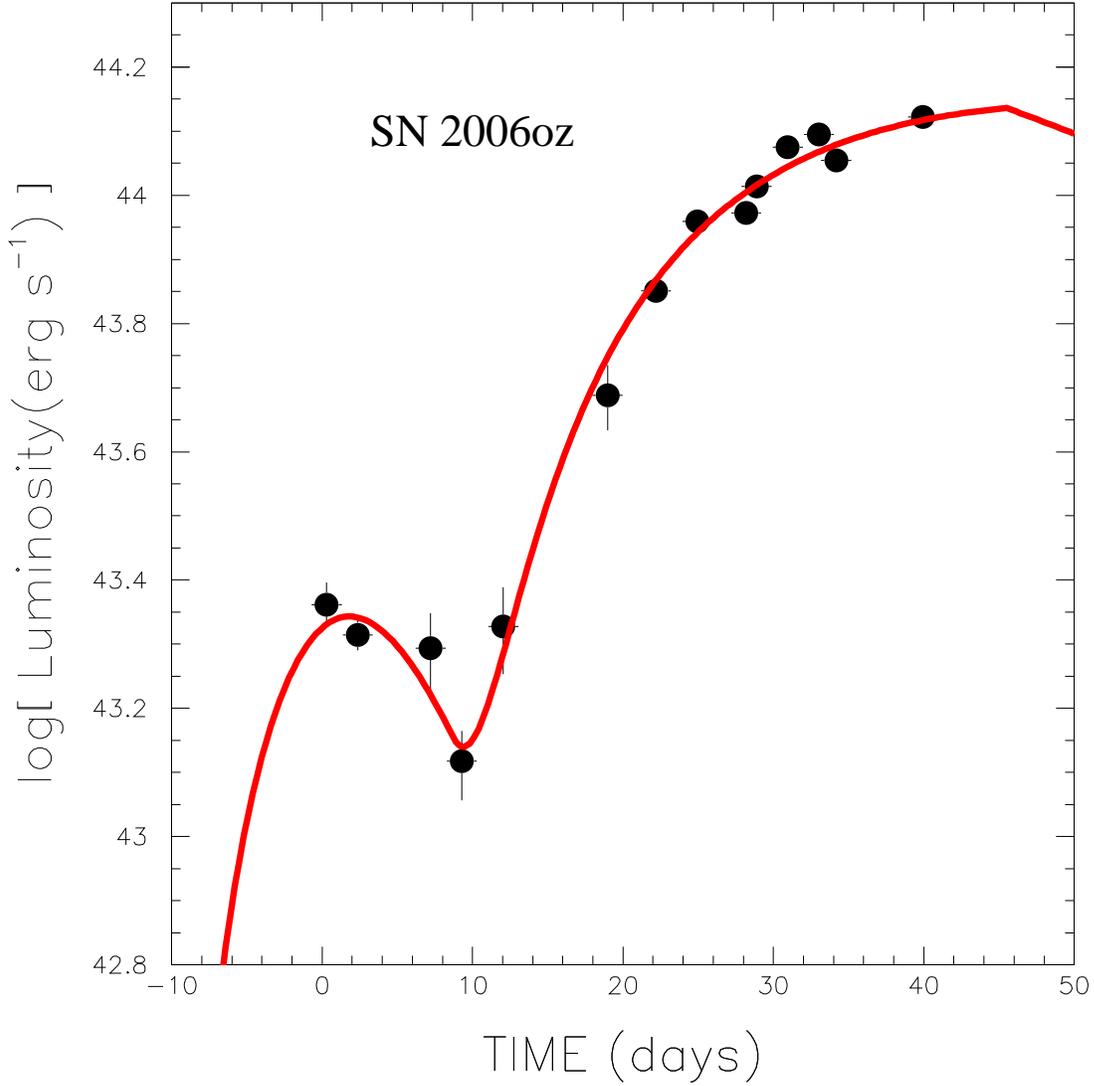,width=16.cm,height=16.cm}
\caption{Comparison between the rest frame light-curve of the bolometric 
luminosity of the superluminous supernova
SN2006oz and a  best fit 
light-curve of a supernova powered by the decay of 0.95  ${M_\odot}$ of
$^{56}$Ni synthesized in the SN explosion followed by  plastic collision 
of the SN debris with a massive shell blown by the 
progenitor star before its explosion.}
\label{Fig2}
\end{figure} 

\newpage 
\begin{figure}[]
\centering
\vspace{-2cm}
\epsfig{file=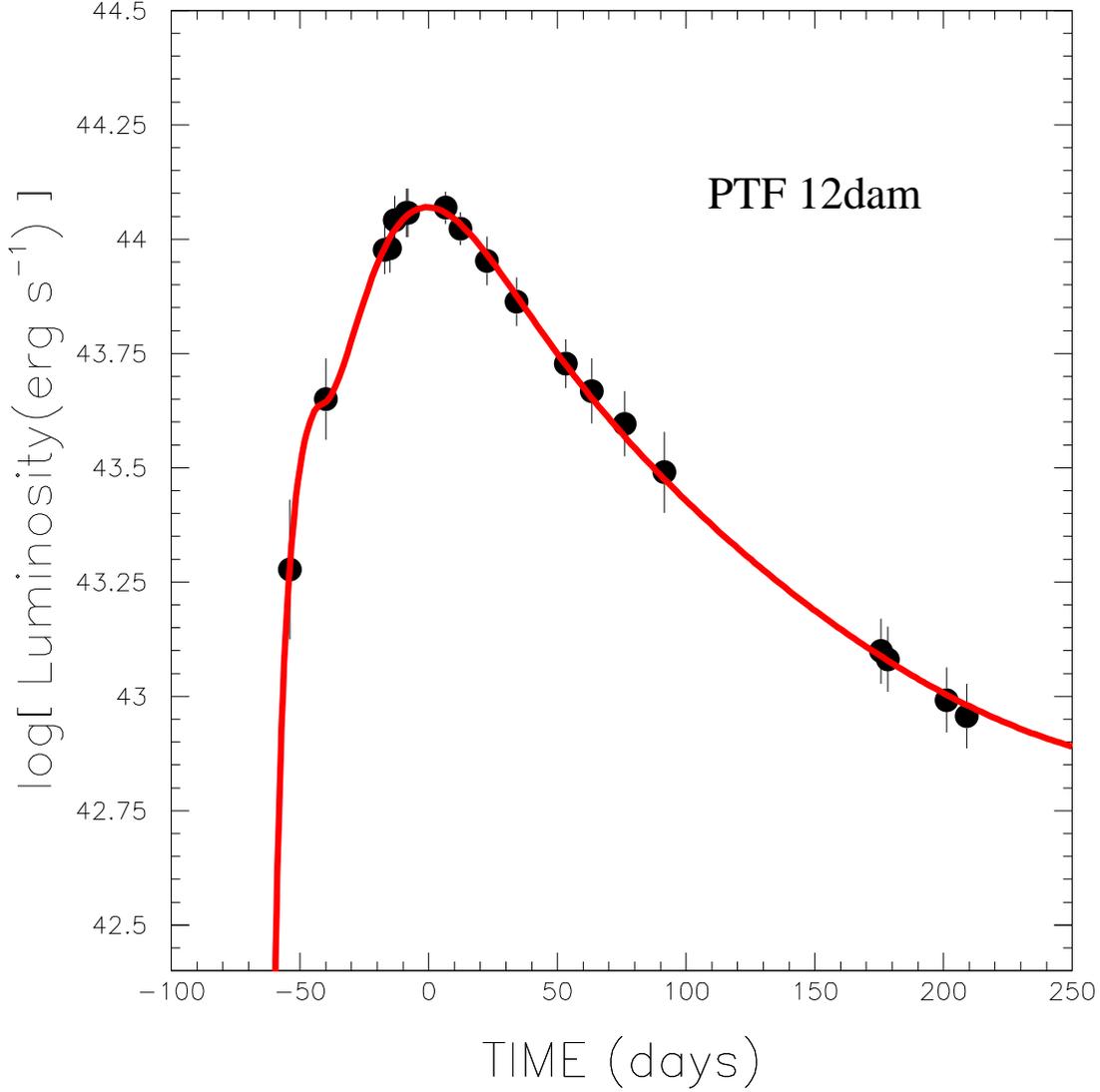,width=16.cm,height=16.cm}
\caption{Comparison of the rest frame light-curve of the bolometric 
luminosity of the superluminous supernova
PTF-12dam (Nicholl et al.~2013) and a  best fit 
light-curve of a supernova powered by the decay of 2.45 ${M_\odot}$ of
$^{56}$Ni synthesized in the SN explosion followed by  plastic collision 
of the SN debris with a massive shell blown by the 
progenitor star before its explosion.}
\label{Fig3}
\end{figure}

\newpage 
\begin{figure}[]
\centering
\vspace{-2cm}
\epsfig{file=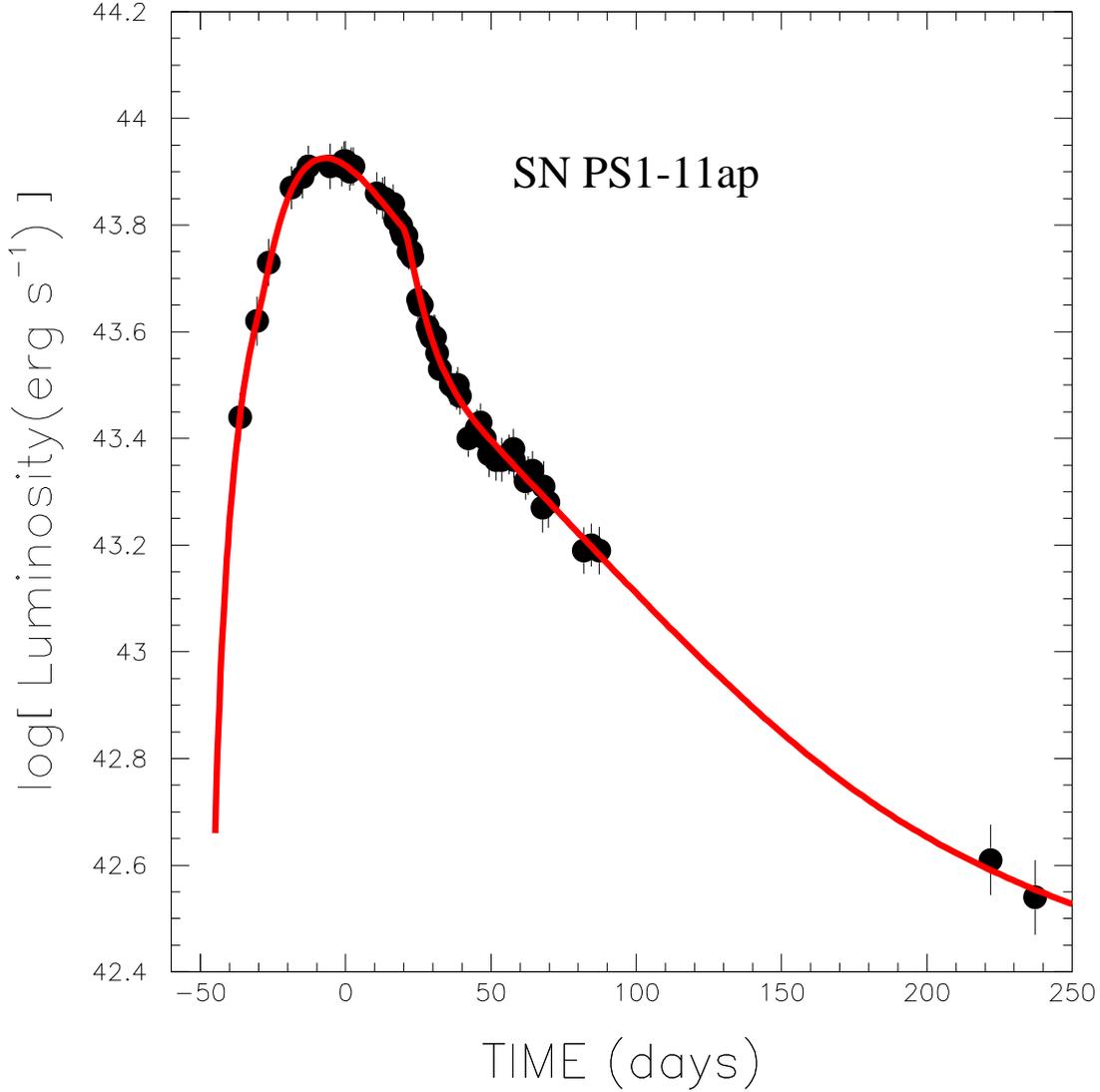,width=16.cm,height=16.cm}
\caption{Comparison between the rest-frame bolometric light-curve of the 
SLSN-Ic ps1-11ap(Nicholl et al.~2013)
and the best fit light-curve of a supernova powered by a plastic
collision of its debris with a massive shell blown by the 
progenitor star before the SN explosion.} 
\label{Fig4}
\end{figure}

\newpage 
\begin{figure}[]
\centering   
\vspace{-2cm}
\epsfig{file=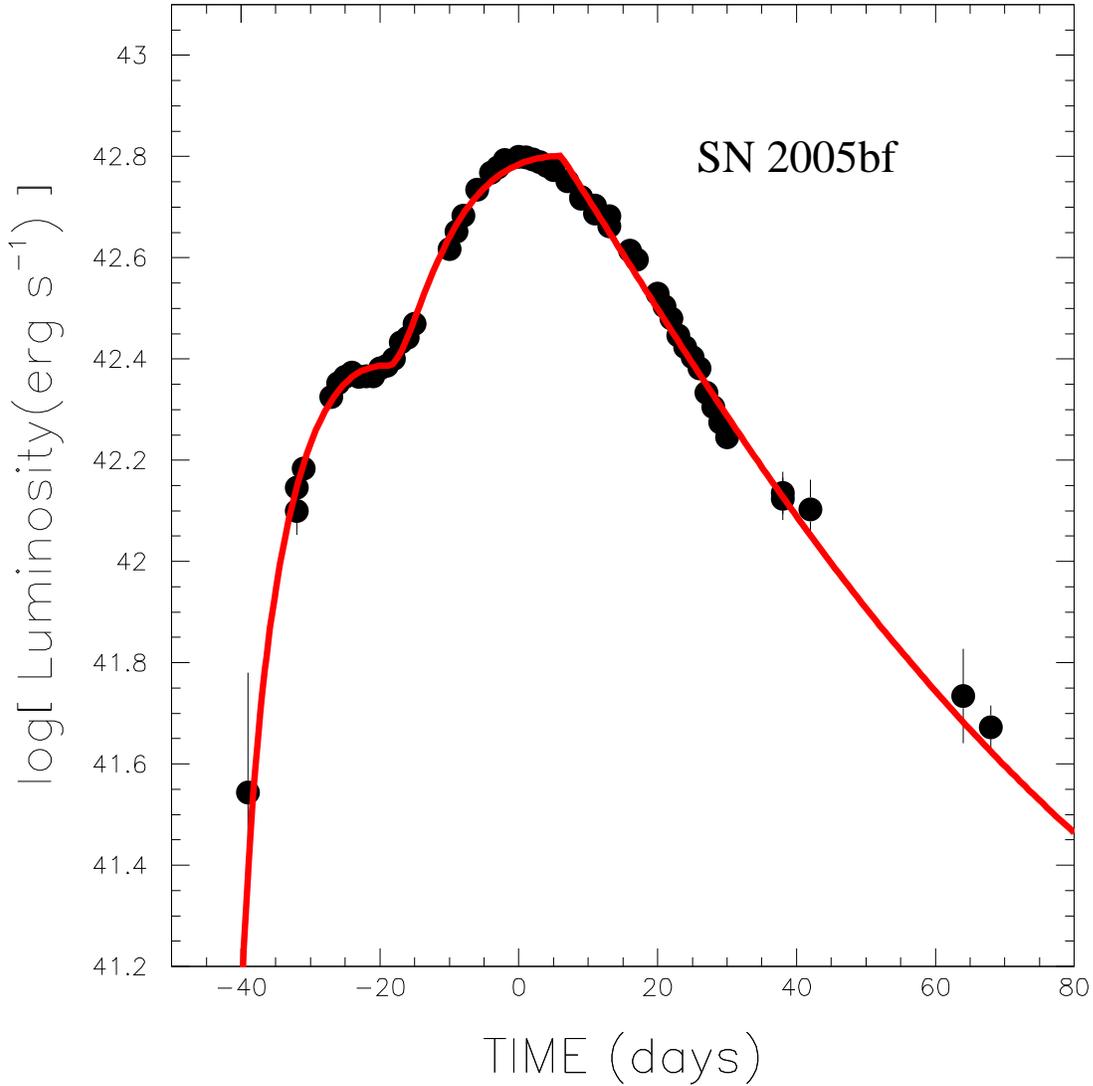,width=16.cm,height=16.cm}
\caption{Comparison between the rest frame bolometric  light curve of the 
interacting 
supernova  SN2005bf (Folatelli et al.~2006) and its best fit bolometric
light-curve powered by the decay of 
${\rm 0.15\, M_\odot}$ of $^{56}$Ni synthesized in the explosion and 
by a subsequent plastic collisions 
of the SN debris with a sucession of massive shell ejected by the 
progenitor star  before its SN
explosion.} 
\label{Fig5}
\end{figure}

\newpage    
\begin{figure}[]
\centering  
\vspace{-2cm}
\epsfig{file=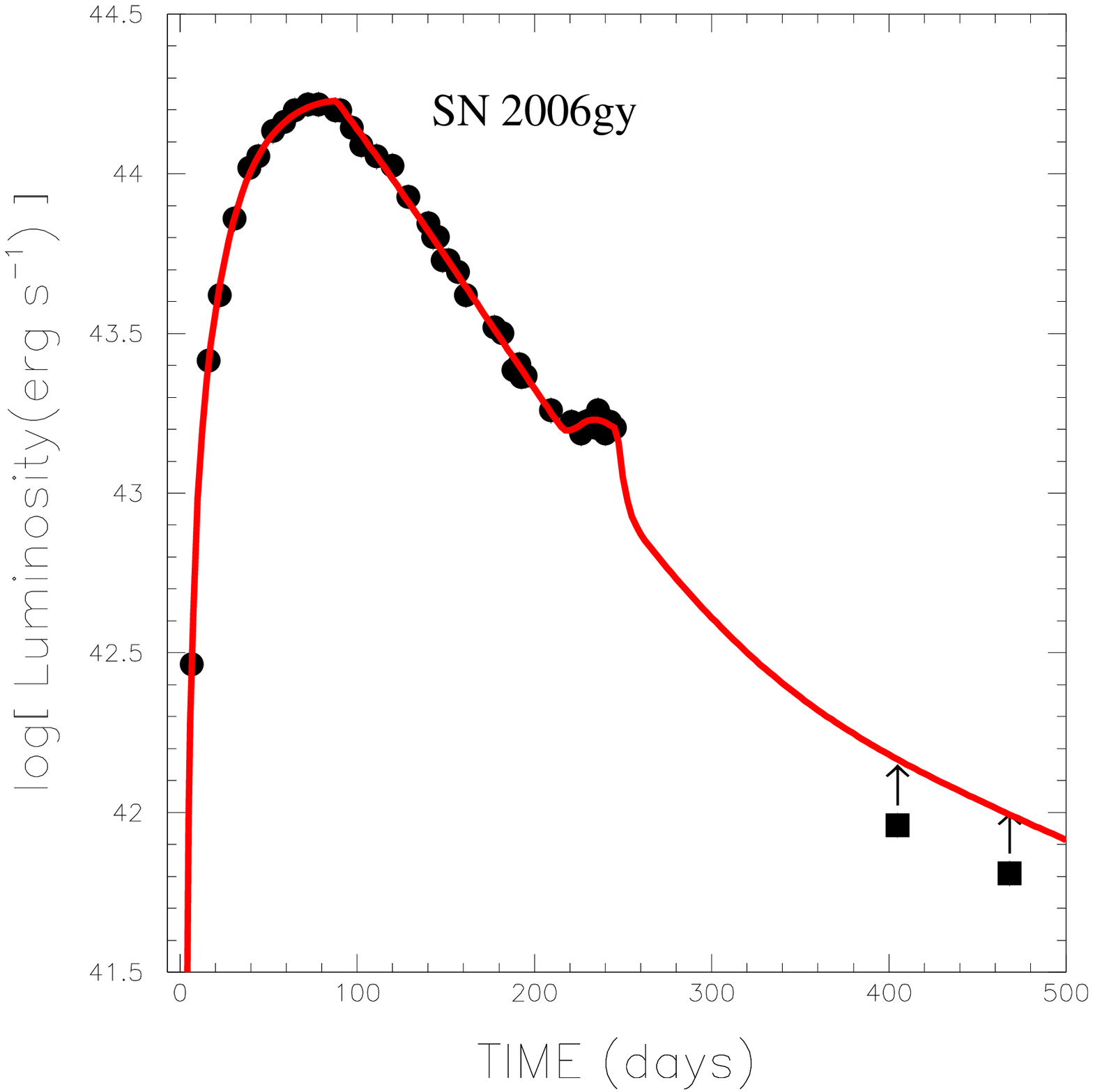,width=16.cm,height=16.cm}
\caption{Comparison between the rest-frame bolometric light-curve of the 
super-luminous  SN20006gy (Smith et al.~2008) and the
best fit light-curve of a supernova dominated  by
two plastic collision of its debris with two  massive shells blown by 
the progenitor star before the SN explosion.}
\label{Fig6}
\end{figure}

\newpage    
\begin{figure}[]
\centering  
\vspace{-2cm}
\epsfig{file=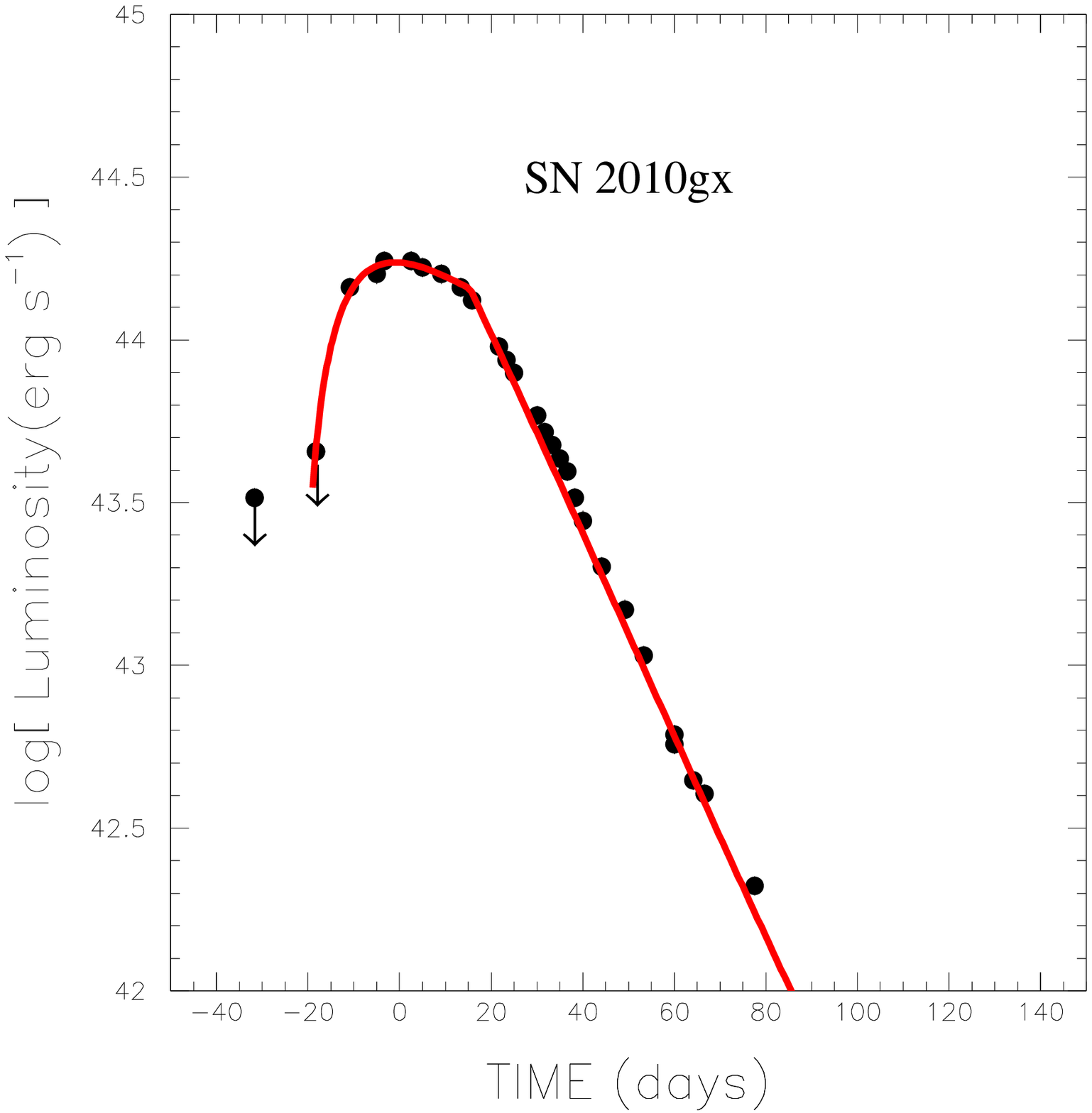,width=16.cm,height=16.cm}
\caption{Comparison between the rest-frame bolometric light-curve of the 
super-luminous  SN2010gx (Pastorello et al.~2010) and the
best fit light-curve of a supernova dominated  by
a plastic collision of its debris with a massive shell blown by 
the progenitor star before the SN explosion.}
\label{Fig7}
\end{figure}


\begin{thebibliography}

\bibitem[Arnett(1982)]{Arnett1982}
Arnett, W. D., 1982, ApJ, 253, 785A

\bibitem[Barkat(1967)]{Barkat1967}
Barkat, Z., Rakavy, G. \& Sack, N. 1967, PRL, 18, 379
 
\bibitem[Chevalier(1982)]{Chevalier1982}
Chevalier, R. A., 1982, ApJ, 258, 790

\bibitem[Colgate and McKee(1992)]{Colgate1992}
Colgate, S. A. \& McKee, C. 1969, ApJ, 157, 623

\bibitem[Colgate et al.(1980)]{Colgate1980}
Colgate, S. A., Petschek, A. G., \& Kriese, J. T. 1980, ApJ, 237, L81

\bibitem[Dado(2009)]{Dado2009}
Dado, S., Dar, A. \& De R\'ujula, A. 2009, ApJ, 696, 994

\bibitem[Dado(2012)]{Dado2012}
Dado, S. \& Dar, A. 2012, arXiv:1207.3630

\bibitem[Dado(2013)]{Dado2013}
Dado, S. \& Dar, A. 2013, arXiv:1301.3333

\bibitem[Dar(2000a)]{Dar2000a}
Dar, A. \&  De R\'jula, A. 2000a, Proc. 
Results and Perspectives in Particle Physics (Ed. Mario 
Vol. 17, p.13 


\bibitem[Dar(2004)]{Dar2004}
Dar, A. \& De R`ujula, A., 2004, Phys. Rept. 405, 203


\bibitem[Drake et al.(2009)]{Drake2009} 
Drake, A. J., et al. 2009, ApJ,696, 870

\bibitem[Folatelli et al.(2006)]{Folatelli2006} 
Folatelli, G., et al. 2006, ApJ, 641, 1039


\bibitem[Gal-Yam(2012)]{Gal-Yam(2012}
Gal-Yam, A., 2012, Science, 337, 927

\bibitem[Grassberg et al.(1971)]{Grassberg1971}
Grassberg, E. K., Imshennik, V. S. \& Nadyozhin, D. K. 
1971, Ap\&SS, 10, 28 

\bibitem[Hadjiyska et al.(2012)]{Hadjiyska2012}
Hadjiyska, E.,  et al. 2012, IAU Symposium, 285, 324

\bibitem[Hubbel(1982)]{Hubbel1982}
Hubbell, J. H. 1982, J. of Applied Rad. and isotopes, 33, 1269

\bibitem[Inserra(2013)]{Inserra2013}
Inserra, C., et al. 2013, arXiv:1307.1791v2 

\bibitem[Kaiser(2010)]{Kaiser2010}
Kaiser, N., et al. 2010, Proc. SPIE, 7733, 12K

\bibitem[Kasen(2010)]{Kasen2010}
Kasen, D., \& Bildsten, L. 2010, ApJ, 717, 245

\bibitem[Kasen(2011)]{Kasen2011}
Kasen, D., Woosley, S. E. \& Heger, A. 2011, ApJ, 734, 102

\bibitem[Katz(1982)]{Katz1982}
Katz,  J.I., 1982, ApJ, 260, 371.

\bibitem[Katz(1982)]{Katz1982}
Katz,  J.I., 2013, arXiv:1307.0586

\bibitem[Kouveliotou(1998)]{Kouveliotou1998}
Kouveliotou, C., et al. 1998, Nature, 393, 235  

\bibitem[Leloudas(2012)]{Leloudas2012}
Leloudas. G., et al. 2012, A\&A, 541, 129

\bibitem[Margutti(2013)]{Margutti2013}
Margutti, R., et al. 2013, arXiv:1306.0038

\bibitem[Marsden(1999)]{Marsden1999}
Marsden, D., Rothschild, R. E. \& Lingenfelter, R. E. 1999, ApJ, 520, L107

\bibitem[Nicholl(2013)]{Nicholl2013}
Nicholl, M.,  et al. 2013, Nature 502, 346 

\bibitem[McCrum(2013)]{McCrum2013}
McCrum, M., et al. 2013, arXiv:1310.4417

\bibitem[Ofek(2007)]{Ofek2007}
Ofek, E. O., et al. 2007, ApJ, 659, L13

\bibitem[Ofek(2013)]{Ofek2013}
Ofek, E. O., et al. 2013, Nature, 494, 65

\bibitem[Pastorello et al.(2010)]{Pastorello2010}
Pastorello, A., et al. 2010, ApJ, 724, L16

\bibitem[Pastorello et al.(2013)]{Pastorello2013}
Pastorello, A., et al. 2013, ApJ, 767

\bibitem[Quimby et al.(2005)]{Quimby2005}
Quimby, R. M., et al. 2005, Bulletin of
the American Astronomical Society, 37, 171.02

\bibitem[Quimby et al.(2006)]{Quimby2006}
Quimby, R., 2006, Central Bureau Electronic Telegrams, 644, 1

\bibitem[Quimby et al.(2007)]{Quimby2007}
Quimby, R. M., et al. 2007, ApJ, 668, L99

\bibitem[Rakavy(1967)]{Rakavy1967}
Rakavy, G. \& Shaviv, G., 1967, ApJ. 148, 803

\bibitem[Rau(2009)]{Rau2009}
Rau, A., et al. 2009, PASP, 121, 1334

\bibitem[Rea(2010)]{Rea2010}
Rea, N., et al. 2010, Science, 330, 944

\bibitem[Rea(2012)]{Rea2012}
Rea, N. et al.  2012, ApJ, 754, 27

\bibitem[Rea(2013)]{Rea2013}
Rea, N. et al. 2013, arXiv:1311.3091 

\bibitem[Smith(2007)]{Smith2007}
Smith, N., et al. 2007, ApJ, 666, 1116

\bibitem[Smith(2007)]{Smith2007}
Smith, N \& McCray R., 2007, ApJ, 671, L17

\bibitem[Smith(2008)]{Smith2008}
Smith, N., et al. 2008, ApJ, 686, 485

\bibitem[Smith(2013)]{Smith2013}
Smith, N., et al. 2013, MNRAS, 434, 2721

\bibitem [Thompson(1995)]{Thompson1995}
Thompson, C. \& Duncan, R. C. 1995, MNRAS, 275, 255

\bibitem [Thompson(1996)]{Thompson1996}
Thompson, C. \& Duncan, R. C.  1996, ApJ, 473, 322
 
\bibitem[Usov(1992)]{Usov1992 }
Usov V. V., 1992, Nature, 357, 472

\bibitem[Woosley(2007)]{Woosley2007} 
Woosley S. E, Blinnikov S \&  Heger A. 2007, Nature, 450, 390

\bibitem[Woosley(2010)]{Woosley2010}
Woosley, S. E.,  2010, ApJ, 719, L204

\bibitem[Zhang(2001)]{Zhang2001}
Zhang B. \&  Meszaros P., 2001, ApJ, 552, L35


\end{thebibliography}
\end{document}